# Electron-phonon coupling and a resonant-like optical observation of a band inversion in topological crystalline insulator $Pb_{1-x}Sn_xSe$

M. Woźny[1], W. Szuszkiewicz[1, 2, *], M. Dyksik[3], M. Motyka[3], A. Szczerbakow[2], W. Bardyszewski[4], T. Story[2, 5], J. Cebulski[1]

[1]Institute of Physics, College of Natural Sciences, University of Rzeszów, S. Pigonia 1, PL-35310 Rzeszów, Poland

[2]Institute of Physics, Polish Academy of Sciences, Aleja Lotników 32/46, PL-02668 Warszawa, Poland

[3]Wrocław University of Science and Technology, Department of Experimental Physics, PL-50370, Wrocław, Poland

[4]Institute of Theoretical Physics, Faculty of Physics, University of Warsaw, Pasteura 5, PL-02093 Warszawa, Poland

[5]International Research Centre MagTop, Institute of Physics, Polish Academy of Sciences, Aleja Lotników 32/46, PL-02668 Warszawa, Poland

**Abstract**

The optical reflectivity of $Pb_{0.865}Sn_{0.135}Se$ and $Pb_{0.75}Sn_{0.25}Se$ solid solutions was measured in the THz spectral region energetically corresponding to bulk optical phonon excitations and in the temperature range from 40 K to 280 K. The analysis of $Pb_{0.75}Sn_{0.25}Se$ data performed within the dynamic dielectric function formalism revealed a new effect due to the electron-phonon coupling resulting in resonant changes of LO phonon frequency for energy gap equal to zero or to LO phonon energy. This effect is absent for $Pb_{0.865}Sn_{0.135}Se$ that exhibits an open energy gap with trivial band ordering at all temperatures. These results show that reflectivity measurements in the THz range constitute a versatile experimental method for precise determination of band inversion in narrow-gap topological materials. For $Pb_{0.75}Sn_{0.25}Se$ the transition from trivial insulator to topological crystalline insulator phase takes place at temperature $T_0 = (172 \pm 2)$ K.

* szusz@ifpan.edu.pl

The studies of topological properties of semiconducting and semimetallic materials and related topological phase transitions attract a lot of attention in recent years being one of the hot topics of contemporary condensed matter physics. The canonical topological materials, like bismuth or antimony chalcogenides, are narrow-gap semiconductors with an inverted band structure, for which electronic Dirac-like metallic states protected by the time-reversal symmetry occur on the surface of insulating bulk crystals [1, 2]. Another class of Dirac topological materials are topological crystalline insulators (TCI) for which crystalline symmetry guarantees the existence of surface Dirac states [3]. It is expected that due to their particular electronic properties topological insulators could serve as a new type of electrical conductors with almost no energy dissipation in nanoscale interconnects, improve presently considered devices in the area of thermoelectricity and infrared or THz optoelectronics as well as constitute materials platform for quantum computation [1, 2].

While considering possible applications or devices taking advantage of the unique electronic properties of TCI materials the crucial point is the determination of topological phase diagram, in particular temperature, pressure and chemical composition corresponding to topological transitions. Several experimental methods were applied in the past for this purpose. The bulk band gap closing at the topological phase transitions followed by the opening afterward is accompanied by the inversion of the symmetry character of the bottom of the conduction band and the top of the valence band. It was the reason that a big part of experiments applied so far for studies of solid solutions which could exhibit the trivial insulator – TCI phase transition are focused on demonstration of the presence of inverted bulk band structure. However, mentioned above important modifications of the band structure usually have a minor impact on a number of electrical and optical properties of investigated material and a clear demonstration of the bulk band inversion is a challenging task.

The $Pb_{1-x}Sn_xSe$ is a TCI material as it was demonstrated a few years ago using the angle-resolved photoemission spectroscopy (ARPES) [4]. This solid solution is one of the best known materials exhibiting the topological transition from a trivial insulator to TCI phase when modifying chemical composition, temperature, or pressure [5-7]. The $Pb_{1-x}Sn_xSe$ solid solutions belongs to the well-known, IV-VI based, narrow-gap semiconductors family. For the chemical composition $0 \leq x \leq 0.37$ it crystallizes in the rock salt structure and the conduction and valence band edges are located at the L points of the Brillouin zone. Below some critical composition $x$ at all temperatures the conduction band is predominantly derived from the cation $6p$ (Pb) or $5p$ (Sn) orbitals ($L_6^-$ symmetry), whereas the highest valence band is in majority formed from the anion $4p$ (Se) orbitals ($L_6^+$ symmetry). With an increasing Sn content at given temperature depending on $x$ (where $0.18 \leq x \leq 0.37$) the system undergoes energy band inversion and the symmetries of these bands above the critical $x$ value change respectively.

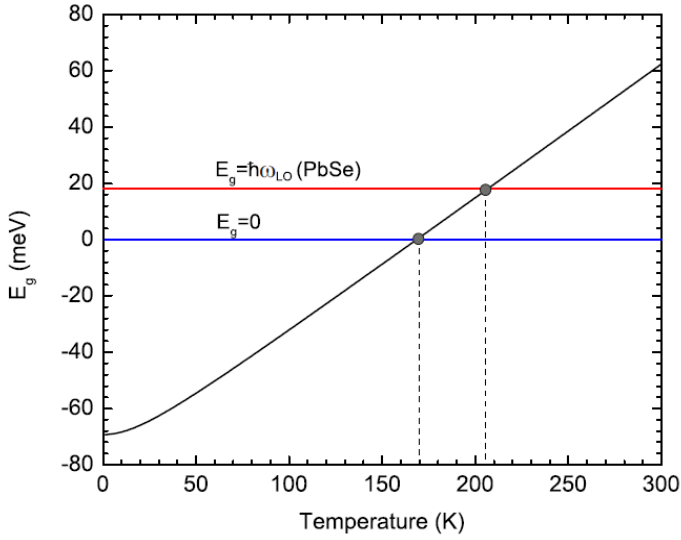

Figure 1. Temperature dependence of the energy gap calculated from Eq. (1) for $Pb_{0.75}Sn_{0.25}Se$ solid solution.

The dependence of $Pb_{1-x}Sn_xSe$ energy gap $E_g$ (in meV) on temperature $T$ and chemical composition $x$ of the solid solution can be described by the expression [8]:

$$Eg(x,T) = 161 - 969x + \sqrt{144 + (0.477T)^2} \quad . \tag{1}$$

The temperature dependence of the energy gap for $Pb_{0.75}Sn_{0.25}Se$ crystal resulting from Eq. (1) shown in Fig. 1 predicts the zero gap and band inversion at $T_0 = 170$ K. The gap is always open for $Pb_{0.185}Sn_{0.135}Se$ crystal – our topologically trivial reference material.

The first experimental method to estimate the temperature of band inversion for $Pb_{1-x}Sn_xSe$ solid solution was based on the observation of a broad minimum in the temperature dependence of resistivity [9] and $T_0 = 150$ K for $x = 0.23$. The authors did not estimate possible experimental error for this value, but it probably could be as high as ± 20 K. The possibility of determination of pressure corresponding to the band inversion for $Pb_{1-x}Sn_xSe$ solid solution was demonstrated by pressure modification of the laser diode emission energy [10] and by pressure-dependent minimum of plasma frequency observed in the reflectivity spectrum [11].

The evidence of the band inversion in $Pb_{1-x}Sn_xSe$ bulk crystals was also obtained from thermoelectric measurements of the Seebeck and the Nernst-Ettingshausen effects [12, 13]. The theoretical analysis of thermoelectric effects in temperature driven band inversion regime pointed out that the Nernst-Ettingshausen effect is particularly strongly affected by the band inversion and reaches maximum when the energy gap vanishes [13]. The $T_0$ value equal to 180 K and to (150 ± 10) K for $Pb_{0.75}Sn_{0.25}Se$ solid solution, respectively, was obtained in these papers. Interestingly, from the analysis of infrared reflectivity data (in the plasma edge region) for different samples with the same solid solution composition ($x = 0.23$) a quite different values of the temperature corresponding to the phase transition, $T_0 = 100$ K and $(160 \pm 15)$ K were proposed in [14, 15], respectively.

The direct evidence of the presence of TCI electronic states for $Pb_{1-x}Sn_xSe$ solid solution obtained by ARPES directly probing the surface electronic structure suggested $T_0$ value close to 100 K [4]. Later on this technique was successfully applied by a few research groups [8, 16 – 18]. The $T_0$ equal to $(130 \pm 15)$ K and $(175 \pm 15)$ K for solid solutions with $x = 0.23$ and 0.27, respectively, was obtained in [8]. The ARPES-data based full composition-temperature ($x$-$T$) topological phase diagram of $Pb_{1-x}Sn_xSe$ crystals was presented in [8, 18]. The TCI states were also detected in scanning tunneling microscopy and spectroscopy (STM/STS) studies locally probing the density of electronic states at the surface [7, 19 – 21]. The ultrahigh mobility surface states observed in $Pb_{1-x}Sn_xSe$ (where $x = 0.23$-$0.25$) by infrared reflectivity in high magnetic fields were reported in [22]. Finally, an important difference in surface phonon dispersion in trivial insulator and TCI phases was also recently demonstrated for $Pb_{0.7}Sn_{0.3}Se$ solid solution using inelastic He atom scattering measurements [23]. The possible additional problem in determination of the trivial insulator-TCI phase transition parameters for a solid solution was pointed out by the density functional theory (DFT) analysis suggesting that this transition is broadened by local chemical (crystal field) disorder with the band inversion transition in a substitutional alloy involving zero-gap states with several cation and anion band crossings (degenerated only for ideal rock salt symmetry) [24]. However, this DFT approach concerns zero temperature limit and neglects the electron-phonon coupling.

In summary, depending on an experimental method, electron conductivity type and free-carrier concentration, published values of determined or estimated topological phase transition temperature for $Pb_{0.77}Sn_{0.23}Se$ solid solution are scattered throughout almost 100 K. Recently, using state of the art Landau level spectroscopy, this temperature was estimated as being close to 150 K in epitaxial layers with the same chemical composition [25]. However, there is still a clear need for a simple and precise experimental tool to determine this transition temperature. In this work, we show that this goal can be achieved by reflectivity measurements and the Kramers-Krönig (KK) analysis in the THz spectral range corresponding to optical phonon frequencies.

The $Pb_{0.75}Sn_{0.25}Se$ and $Pb_{0.865}Sn_{0.135}Se$ solid solution single crystals used in this study were grown by the self-selecting vapor-growth method (SSVG) [26, 27]. Their chemical composition and homogeneity of the crystal was determined by energy dispersive X-ray spectroscopy (for details, see [4, 12]). The 2 mm thick crystal plates with the dimensions 4 x 3 mm$^2$ selected for optical studies were cleaved by the razor blade along (001) natural cleavage plane. The electron transport measurements carried out previously for this $Pb_{1-x}Sn_xSe$ solid solution demonstrated an *n*-type conductivity, for the sample with $x = 0.25$ the electron concentration equal to $3.1 \cdot 10^{18}$ cm$^{-3}$ and $2.4 \cdot 10^{18}$ cm$^{-3}$ at $T = 4.2$ K and $T = 295$ K, respectively [12].

The reflectance measurements were performed with Bruker FTIR vacuum spectrometer Vertex 80v operating in a rapid-scan mode [28]. As a light source the mercury lamp with mylar multilayer beamsplitter was used. The Si bolometer with 700 cm$^{-1}$ cut-off low-pass filter employed as a detector spectrally covered the range of THz frequencies. The measurements were performed from 50 cm$^{-1}$ to 450 cm$^{-1}$ for oblique incidence of 12 degrees utilizing the transmission/reflection unit and temperature changes (40 K – 280 K) provided by the optical cryostat equipped with polyethylene window. The spectral resolution was equal to 1 cm$^{-1}$.

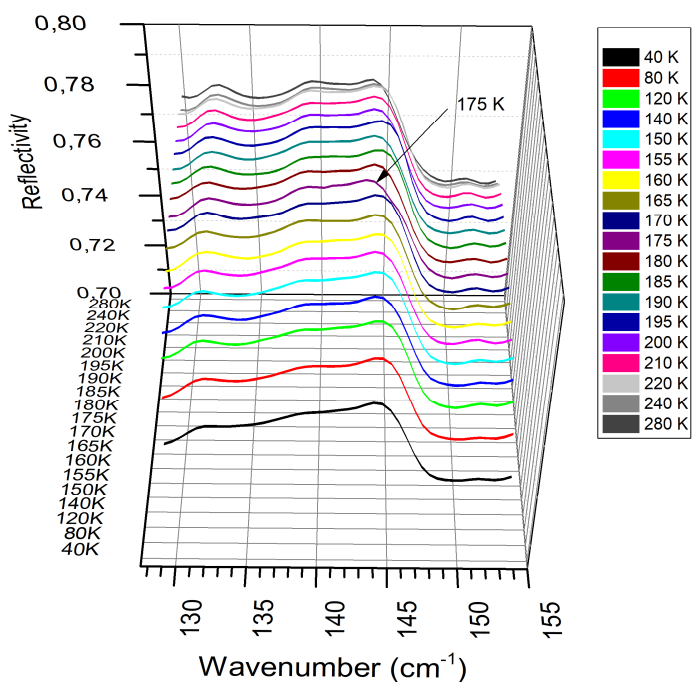

Figure 2. The part of reflectivity curves determined at several temperatures for Pb$_{0.75}$Sn$_{0.25}$Se solid solution. The selected spectral range close to LO phonon frequency is shown.

When analyzing experimental data a well pronounced step on the reflectivity curves was found at about 145 cm$^{-1}$ (Fig. 2). From the measured reflectivity coefficient a real and an imaginary part of dynamic dielectric function (DDF) was calculated using the standard KK relations, the previously described procedure valid for solid solutions was applied [29]. Fig. 3 presents the imaginary part of DDF resulting from the KK analysis. The full width at half maximum (FWHM) of the structure is of the order of 3 cm$^{-1}$, the amplitude passes by a maximum below about 200 K. Its frequency is fixed within 1% but also exhibits a non-monotonous evolution, shown in detail in Fig. 4. The error of experimental data does not exceed 0.12 cm$^{-1}$, the dashed line is a guide for the eye.

The step in reflectivity spectra (Fig. 2) and relevant structure shown in Fig. 3 and analyzed in Fig. 4 are related to the Pb$_{0.75}$Sn$_{0.25}$Se lattice dynamics. The simple Drude-Lorentz model satisfactorily describing previous reflectivity and transmission spectra in the infrared assumed independent damping constants of free-carrier plasma and phonon excitations [10, 13, 14, 30]. However, a deviation from this model can be expected when the light frequency approaches the optical phonon frequency. A strong mixing of the electron plasma and polar phonon modes resulting in the collective motion of electrons and ions causes a significant modification of the effective scattering of electrons by impurities [31]. The step presently observed in reflectivity spectra results from an influence of plasmon-phonon coupling on the electron-impurity interaction in the free carrier absorption and corresponds to the LO phonon mode. This effect was previously observed and theoretically demonstrated for another narrow-gap semiconductor with an inverted band structure, HgSe [32]. The LO frequency equal to about 140 cm$^{-1}$ is expected to be almost the same for PbSe and Pb$_{0.75}$Sn$_{0.25}$Se (see [33]).

The influence of the phonon system on the electron band structure of solids resulting from the electron-phonon coupling was investigated for many years and today is a well-established phenomenon. The opposite effect, i.e. possible influence of the electron system and, in particular, of the band structure on the phonon system was much less explored in the past. The latter influence resulting in a correction to the TO phonon frequency was theoretically investigated for PbTe, material closely related to the Pb$_{1-x}$Sn$_x$Se, providing the following formula for renormalized TO phonon frequency [34] :

$$\omega_T^{*2} = \omega_T^2 - \frac{4\Xi_{cv}^2}{Ma^2 W} \ln \frac{W}{2E_F + E_g} \qquad (2)$$

where $\Xi_{cv}$ is the optical deformation potential matrix element between valence and conduction band, $M$ is the reduced mass of two different ions, $a$ is the lattice parameter, $W$ is the sum of the width of the conduction and the valence band, $E_F$ is the Fermi energy calculated from band edge, and $E_g$ is the energy gap. A small modification of this formula, predicting two discontinuities in the temperature dependence of phonon frequency for $E_g = 0$ and $E_g = \hbar\omega_{TO}$ was shown to be valid for II-VI semiconductor with zero-energy gap [35]. The direct experimental evidences of this effect were demonstrated for Hg$_{1-x}$Cd$_x$Te [35] and Hg$_{1-x}$Zn$_x$Te [36] solid solutions.

The optical phonon softening accompanied by an anomaly in the phonon linewidth resulting from electron-phonon interaction observed in Raman spectra for Sb$_2$Se$_3$ bulk crystal were explained by an electronic topological transition taking place under pressure for this compound [37]. Possible evidence of the electronic band inversion by an analysis of the bulk optical phonon linewidth was also demonstrated in [38]. The presence of a week effect on this phonon frequency was suggested in the Supplemental Material to this paper. In spite of a small value of the latter effect it can be experimentally demonstrated, as it is shown by the present data.

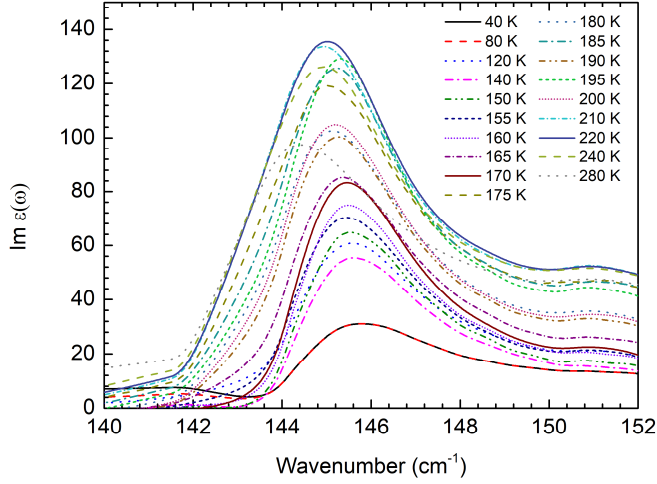

Fig. 3. The imaginary part of DDF calculated from the experimental spectra shown in Fig. 2 according to the procedure described in the text.

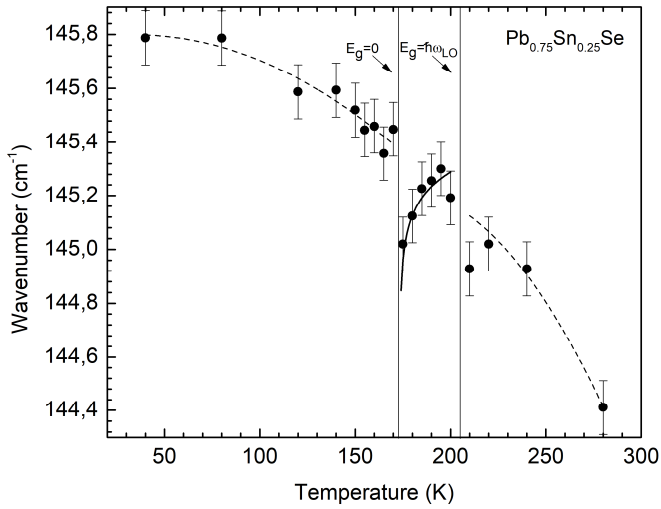

Fig. 4. Temperature evolution of LO phonon frequency resulting from data presented in Fig. 2 and Fig. 3 (details in text).

The effect analogous to that demonstrated in [35, 36] can be expected for LO phonon in $Pb_{1-x}Sn_xSe$ solid solution. The first, clear discontinuity in a temperature evolution of LO phonon frequency seen in Fig. 4 at temperature $T = 172 \pm 2$ K corresponds to expected energy gap value $E_g = 0$ [8]. It can be considered as a new signature of temperature-induced band inversion in the bulk corresponding to the trivial insulator – TCI phase transition. The second, less pronounced anomaly appears at temperature for which $E_g = \hbar\omega_{LO}$. Moreover, a form of the LO frequency temperature evolution from about 170 K to 205 K can be perfectly described by a formula analogous to that previously applied for the TO phonon mode. The only required modifications are a replacement of TO by LO frequency and some reduction of the deformation potential in Eq. (2). The solid line in Fig. 4 was calculated according modified in this way formula using the following set of parameters: $W = 18$ eV, $E_F = 9$ meV, $a = 6.12$ Å, $M = 9.49 \cdot 10^{-26}$ kg, the optical deformation potential $\Xi_{cv} = 0.5$ eV. The value of $E_F = 23$ meV follows from known electron concentration $n = 2.4 \cdot 10^{18}$ cm$^{-3}$ and Eq. (2) from Ref. [13]. It could be overestimated due to possible modification of the form of band edge. The hybridization-driven modification of energy gap and some flattening of the energy dispersion in Pb$_{1-x}$Sn$_x$Se were indeed previously suggested [14].

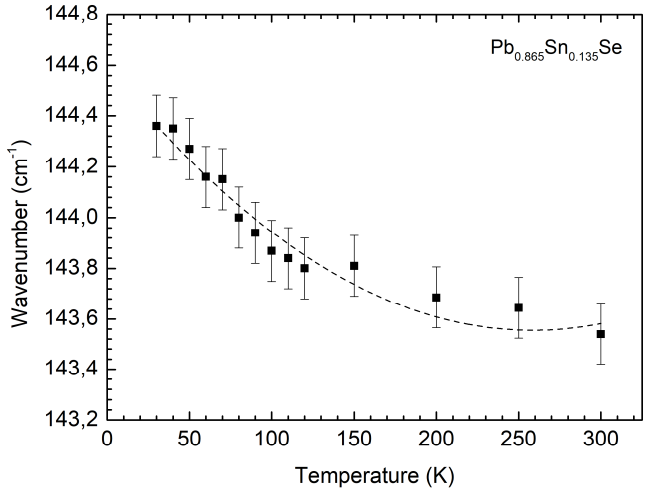

Fig. 5. Temperature evolution of LO phonon frequency for Pb$_{0.865}$Sn$_{0.135}$Se solid solution crystal determined by the same procedure as that applied for Pb$_{0.75}$Sn$_{0.25}$Se sample.

The reflectivity spectra similar to those determined for the Pb$_{0.75}$Sn$_{0.25}$Se sample were measured for the comparison for Pb$_{0.865}$Sn$_{0.135}$Se single crystal (Fig. 5). According to the Eq. (1) the solid solution composition for $x = 0.135$ corresponds to the trivial band ordering at all temperatures. One can't expect an anomaly in the temperature dependence of LO phonon mode for this sample. Nevertheless, the reflectivity spectra were measured at temperatures T < 130 K every 10 K. This experimentally determined dependence is shown in Fig. 5. As previously, the dashed line is a guide for the eye. According to the prediction the observed smooth temperature dependence of LO phonon frequency at low temperatures excludes possible change of the band ordering for the Pb$_{1-x}$Sn$_x$Se solid solution crystal with $x = 0.135$.

All proposed up to date experimental methods to detect the TCI phase or to determine parameters of possible phase transition from a trivial insulator to a TCI phase have some disadvantages or at least their applications are limited to some extent. The ARPES measurements can be performed on a non-oxidized, clean crystal surface only. In the case of bulk samples this technique requires single crystals of a big size (several mm long) because of a necessity of their cleavage under very high vacuum conditions. It could be difficult to satisfy such condition for many materials. The experimental methods taking advantage of electron transport techniques requires first a careful preparation of high-quality electric contacts what could depend on the electron concentration and could be different for n-type and p-type samples. In the case of new or less-known materials it could be a real difficulty hampering an application of electron transport techniques. The published optical method based on a modification of absorption edge slope results from the analysis of reflectivity in the mid-infrared spectral range versus temperature or from direct measurements performed for a few micrometer thin plane-parallel slab. A sensitivity of the first method is not very high due to very limited influence of this absorption mechanism on the reflectivity coefficient. A careful analysis of the absorption edge requires transmission measurements with the use of very thin (a few micrometer thick) samples and not all materials can be grown in the form of such thin plates. In principle, a preparation of required for transmission measurements sample starting from a bulk crystal should be possible [39]. However, it is really a challenging experiment not applicable for a practical use. Importantly, none of the experimental methods previously applied has a resonance-like character. In most of cases the temperature of the trivial insulator-TCI phase transition was previously estimated from a modification of the slope of a smooth curve (methods based on the resistivity measurements or on the analysis of a reflectivity). A part of methods based on the maximum or the minimum value of selected parameter (the Fermi energy, the plasma frequency, the electron effective mass) requires complex interpretation of experimental data with several additional assumptions or approximations. Probably the best up to date experimental determination of the phase transition temperature based on a sophisticated analysis of the high field magnetospectroscopy data [25] is too complicated and time consuming method for a practical, 'every-day' use.

The presently proposed method of determination of the band inversion and accompanying phase transition temperature for a solid solution is based on a new effect resulting from electron-phonon interaction. It is free from several limitations listed above and seems to be 'a user friendly' method. The reflectivity measurements do not require a big size of analyzed bulk samples. A typical surface of investigated single crystal plate could be small because the light is focused approximately into a 2 mm diameter spot on the sample surface. This condition can be easily satisfied for a variety of materials. The proposed method is not destructive and the sample surface does not need to meet particular requirements. A natural, cleaved in air, mechanically polished or chemically etched crystal surface is suitable for such kind of measurements. The interpretation of reflectivity curves requires standard, well established KK relations. Last but not least, the anomaly in the temperature dependence of LO phonon mode frequency has a resonant-like character and its energy position can be determined relatively easy. The Fourier spectrometer dedicated to optical measurements in the infrared is a typical laboratory equipment and its experimental possibilities can be easily extended to the reflectivity measurements in the THz spectral range. Taking into account all circumstances mentioned above the proposed method can be an excellent tool to study semiconductors considered as possible candidates for topological materials.

**Acknowledgements**

Valuable comments and discussions with K. Dybko are greatly appreciated. This work was partly supported by National Science Centre (Poland) projects 2014/13/B/ST3/04393 and 2014/15/B/ST3/03833 as well as by the Foundation for Polish Science through the IRA Programme co-financed by EU within SG OP.